%% file: paper.tex
\documentclass[copyright,creativecommons]{eptcs}

\providecommand{\event}{FORECAST 2016}

\usepackage{amsmath,
	    amsfonts,
	    amstext,
	    amssymb,
	    stmaryrd,
	    latexsym,
	    comment,
	    graphicx}

\input{commands}

\input{environments}

\begin{document}

\title{A Formal Framework for Modeling Trust and Reputation in Collective Adaptive Systems}

\author{Alessandro Aldini 
\institute{Dipartimento di Scienze Pure e Applicate, Universit\`a di Urbino,
Urbino, Italy}
\email{alessandro.aldini@uniurb.it}}

\def\titlerunning{A Formal Framework for Modeling Trust and Reputation in CASs}
\def\authorrunning{A.~Aldini}

\maketitle

\begin{abstract} 
Trust and reputation models for distributed, collaborative systems have been
studied and applied in several domains, in order to stimulate cooperation while
preventing selfish and malicious behaviors. Nonetheless, such models have
received less attention in the process of specifying and analyzing formally the
functionalities of the systems mentioned above. The objective of this paper is
to define a process algebraic framework for the modeling of systems that use
$(i)$ trust and reputation to govern the interactions among nodes, and $(ii)$
communication models characterized by a high level of adaptiveness and flexibility.
Hence, we propose a formalism for verifying, through model checking techniques, the
robustness of these systems with respect to the typical attacks conducted
against webs of trust.
\end{abstract}

\section{Introduction}\label{sec:intro}

Trust and reputation management systems~\cite{Jos} can improve the reliability
of the interactions and the attitude to cooperation for several types of
collaborative systems, in various different domains, such as participatory
sensing systems, wireless sensor networks, peer-to-peer services, mobile ad-hoc
networks, user-centric networks, supply networks, and, last but not least,
collective adaptive systems.  Typically, the models proposed for these systems
rely on distributed notions of trust and reputation.  More precisely, trust
management is distributed over all the nodes, which may collaborate with each
others in order to exchange and aggregate personal opinions, calculate trust
scores of target nodes, and disseminate such
values~\cite{eps365559,Yaich:2012:AST:2245276.2232112,CSC11,Momani2010}.  For
instance, trustworthy sensor networks base their ability to collectively process
sensed data on decentralized reputation
systems~\cite{Ganeriwal2008,YuLiZL,Han2014602,NordioInfocom2015,Mousa2015}.
Nodes monitoring the behavior of neighbor nodes in the network maintain
reputation for such nodes. Hence, collaboration among nodes with high reputation
can be strengthened while malicious nodes are excluded from the community, thus
favoring activities like, e.g., intrusion detection, participatory sensing, and
many more.

A web of trust can be established according to a geographical notion of
\textit{group} of nodes, as in crowdsourcing and sensor
networks~\cite{Ganeriwal2008}, or by following community based models, as in
social networks and P2P environments~\cite{ZLH}.  Trust derives from local,
direct observations, e.g., through watchdog mechanisms, quantitatively
represented by scores assigned to rate the result of interactions, and from
second hand information, represented by recommendations provided to a node by
the other nodes of its web of trust. All these values are combined by the
specific trust system to derive, e.g., a computational notion of trust, which is
then used as a belief level to predict either statistically or deterministically
the future behavior of the various network members. 

\begin{example}
In several trust models~\cite{Beth94valuationof,NgaiLyn94,ZLH}, the trust value
of peer $A$ towards peer $C$ through peer $B$ is expressed by a formula of
the form: \[1 - (1 - t_{BC})^{t_{AB}} \] where $t_{IJ}$ is the trust from $I$ to
$J$. Hence, $t_{BC}$ plays the role of a recommendation given to $A$, which is
weighted by the direct trust from $A$ to $B$.
Inspired by this model, in~\cite{ZLH} a notion of \textit{club} is used to aggregate 
multiple self-organizing peers with common needs/features in order to improve 
the efficiency of service discovery/delivery in peer-to-peer collaborative networks. 
Each club includes a special node, called CDSR, with management tasks.
Then, trust is generalized to express relations among clubs. For instance, the
trust from club $X$ to club $Y$, reporting the result of direct experiences
among peers belonging to the two clubs, depends on the amount of positive
experiences $p$ and negative experiences $n$ observed by peers in $X$ when
interacting with peers in $Y$:
\begin{equation}
\label{eq:clubtrust}
t_{XY}(p,n) = \left \{
\begin{array}{ll}
1 - \lambda^{p-n} & \mathit{if\ } p > n \\
0 & \mathit{otherwise}
\end{array}
\right .
\end{equation}
where the configuration parameter $\lambda$ is the probability of reliability
with a single interaction. Instead, the reputation of peer $K \in Y$ as
perceived by the other peers of $Y$ is non-zero only if all the interactions of
such peers with $K$ are positive and depends on the amount $p$ of these direct 
experiences: 
\begin{equation}
\label{eq:trustinclub}
t_{YK} = 1 - \lambda^p. 
\end{equation}

By combining these trust values, we obtain the trust of any peer in club $X$
towards peer $K$ belonging to another club $Y$:
\[
t_{XK} = 1 - ( 1 - t_{YK} )^{t_{XY}}.
\]
\end{example}

\begin{example}

In reputation-based sensor networks~\cite{Ganeriwal2008,LiKato08}, the local, direct
trust from node $I$ to node $J$ is maintained by using a watchdog mechanism in
$I$ reporting the result of each direct experience with $J$. Such a feedback,
which may consist of scores or, more simply, the amount of good behaviors and
of misbehaviors observed, is then used to parameterize a trust formula relying on a
standard Bayesian approach. The calculated trust value thus represents the
expectation estimating the belief level that one node has on another node for a
specific action.
Second hand information can be asked from neighbor nodes, in the form of
recommended trust values reported by such nodes and scaled by a factor
proportional to the trust towards such recommending nodes.

\end{example}

\begin{example}

EigenTrust~\cite{KSG} is a trust system originally proposed for P2P file sharing
systems. Peers rate with value $1$ (resp., $-1$) each satisfactory (resp.,
unsatisfactory) interaction. The local trust $s_{ij}$ from $i$ to $j$ is
computed by summing up the scores of the individual transactions conducted by
peer $i$ with peer $j$. Then, $s_{ij}$ is normalized with respect to $\sum_j
s_{ij}$ in such a way to obtain a trust value $c_{ij}$ between $0$ and $1$, with
$\sum_J c_{IJ} = 1$. These trust values are then aggregated to form a
distributed notion of reputation.
The principle behind the computation of the global trust $t_{ij}$ from $i$ to
$j$ is to combine the opinions of $i$'s neighbors, as follows:
\[\sum_k c_{ik} c_{kj} \]
In matrix notation, given $C$ the matrix $[c_{ij}]$ of all the trust values and
$c_i$ the vector containing the values $c_{ij}$, then the vector $t_i$ of the
values $t_{ij}$ is computed as $C^T \cdot c_i$. Such a mechanism can be iterated by
aggregating the opinions of communities in cascade, i.e., by computing $(C^T)^n
\cdot c_i$. For $n$ large enough, the result converges to the same trust vector for
every peer $i$ in the network, which thus represents the vector of global trust
values.

In PeerTrust~\cite{Xiong2004}, developed for distributed systems, trust towards 
a peer $i$ depends on the amount of known interactions between $i$ and other peers, 
the known feedback reported by such peers, the credibility of such peers, and an adaptive
community context factor for peer $i$. In turn, credibility of a peer $j$ from
the viewpoint of a peer $k$ depends on the recommendations about $j$ provided by
peers that previously interacted with both $k$ and $j$.

In all these examples, the trust-based selection is based on the rule $t_{ij}
\ge \mathit{th}_{i}$, where the trust threshold value $\mathit{th}_{i}$ may
depend on several factors influencing $i$, such as the dispositional trust of
$i$, which represents the initial willingness of the peer $i$ to cooperate with 
unknown peers.

\end{example}

Systems such as those mentioned above are typically verified through
simulation~\cite{Kim2009,Ganeriwal2008,ZLH,KSG} or game theory~\cite{LS}, possibly 
leading to results validating the trust model against attacks like, e.g.:
\begin{itemize}
\item bad mouthing: negative feedback reported by an adversary about the behavior
of a trusted agent;
\item ballot stuffing: positive feedback reported by an adversary about the
behavior of a malicious agent;
\item collusion: attack conducted by multiple adversaries which act together
with the aim of damaging a honest agent;
\item on-off: attack conducted by an adversary alternating between normal
behaviors and misbehaviors.
\item sybil: attack conducted by an adversary generating multiple identities
with the aim of flooding the system with fake information or misbehaviors.
\item white-washing: attack conducted by a misbehaving adversary who leaves the system
whenever her reputation is compromised and then rejoins it using a different
identity.
\end{itemize}
However, the lack of formal validation can be seen as a weakness, especially in
such a complex framework in which attacks and countermeasures depend on the
flexibility and on the dynamic behavior of the web of trust~\cite{YuLiZL,MP}.
Classical verification techniques, like model checking, have demonstrated their
adequacy in the validation process of systems with respect to properties like safety, 
reliability, security, and performance. On the other hand, they have not
received the same attention in the setting of trust and reputation (see, e.g.,
\cite{Ald15} and the references therein). To cite few representative examples in the
setting of model checking based analysis, Reith et al.~\cite{Reith2007} verify
delegation mechanisms in access control, which can be viewed as a form of trust
management, while He et al.~\cite{He2010} apply the same approach to the
verification of chains of trust. Finally, in~\cite{AB14,KPS} the PRISM model
checker is used to estimate the tradeoff between trust-based incentives and
remuneration-based incentives in cooperative user-centric networks.

In this paper, which is inspired by~\cite{Ald15}, we present a process algebraic
framework for the modeling and, therefore, analysis of trust-based adaptive
systems. With respect to~\cite{Ald15}, the proposed framework offers different
ways of modeling trust and trust-based choices, and introduces mobility and
collaboration aspects affecting the establishment and management of dynamic and
adaptive webs of trust.  To this aim, a notion of environment is modeled
explicitly that guides the communication and, as a consequence, the trust
relationships, among dynamic agents. Historically, starting with the Ambient
Calculus~\cite{CardelliGordon}, and until the most recent
proposals~\cite{EPTCS194.2,MELA}, several process calculi have been defined that
represent mobile computation with a notion of environment. With respect to such
proposals, the contribution of this paper is a dynamic communication model
relying on trust relationships.

The rest of the paper is organized as follows. In Section~\ref{sect:model}, we
present the formal framework for the description of an agent-based network of
trust. We first define a basic calculus of sequential processes and then we show
how to model communications based on trust relations. Then, in
Section~\ref{sect:example} we show the adequacy of such a framework by
presenting two real-world examples. In Section~\ref{sect:mc}, we briefly discuss
how to model check trust-based properties and, finally, in
Section~\ref{sect:conc} we comment on future directions for the proposed approach.

\section{Modeling an agent-based web of trust}\label{sect:model}

All the examples shown in the previous section emphasize that the ingredients needed
to feed a trust model for distributed, adaptive systems are:
\begin{enumerate}
\item the set of direct experiences affecting a local notion of trust. A direct
experience is expressed quantitatively by a positive/negative score assigned to
evaluate an interaction. 
\item the set of groups of agents collaborating, e.g., through the exchange of 
recommendations, in order to calculate a global notion of trust. It is worth
observing that the composition of such groups may be characterized by high
levels of flexibility. 
\end{enumerate}

It is worth observing that in the following we abstract from the way in which the 
basic parameters concerned with local and global notions of trust are combined to
compute opinions governing the decision making process, which is a task specific
of the trust model adopted. Instead, we concentrate on the specification of the
behavior of agents and on the establishment of their networks of trust. For this
purpose, as we will see, in the semantics of our formal specification language
we have rules describing $(i)$ how the basic parameters needed by the trust
system are calculated and maintained, and $(ii)$ how the results computed by the
trust system, i.e., the $t_{IJ}$ values, are then used to govern the trust-based
interactions. All the machinery taking in input the basic parameters mentioned
above and returning as output the trust values is hidden and left to the
specification of the trust model.

Moreover, to simplify the presentation, unless differently specified we restrict
our consideration to systems in which one type of service is provided within the
network. In order to generalize, it is sufficient to replicate as many instances of
the trust infrastructure as the number of different services modeled in the
system, because trust-based beliefs are specific to the required service.

\subsection{Basic Calculus}

We denote with $\mathit{Name}$ the set of visible action names, ranged over by
$a, b, \ldots$, and we assume that $\mathit{Name} = \mathit{Name}_o \cup
\mathit{Name}_i$, where $\mathit{Name}_o$ and $\mathit{Name}_i$ are disjoint and
represent the sets of output actions and input actions, respectively. The 
fresh name $\tau$ is used to represent invisible, internal actions. 
We also use $\alpha, \ldots$ to express visible and internal actions.

The set of terms of the basic calculus for sequential processes is generated
through the following syntax:
\[\begin{array}{l}
P \; ::= \; \nil \mid \alpha \, . \, P \mid P + P \mid B 
\end{array}\] 
where we have the constant $\nil$ denoting the inactive process, the classical
algebraic operators for prefix and nondeterministic choice, and a constant based
mechanism for expressing recursive processes. As usual, we consider only guarded
and closed process terms. 

Then, the semantics of process terms is expressed in terms of labeled transition
systems.

\begin{definition}\label{lts}
A labeled transition system (LTS) is a tuple $(Q,q_0,L,R)$, where $Q$ is a
finite set of states (with $q_0$ the initial one), $L$ is a finite set of
labels, and $R \subseteq Q \times L \times Q$ is a finitely-branching transition
relation.
\end{definition}
		
In the following, $(q,l,q') \in R$ is denoted by $q \arrow{l} q'$. Then, the
behavior of process term $P$ is defined formally by the smallest LTS
$(Q,q_0,L,R)$ such that $Q$ is the set of process terms of our basic calculus
(with $P$ representing the initial state $q_0$), $L = \{\tau\} \cup
\mathit{Name}$, and the transitions in $R$ are obtained through the application
of the operational semantics rules of Table~\ref{tab:sem}. The semantics of
process term $P$ is denoted by $\lsp P \rsp$.

\begin{table}
\caption{Semantics rules of the basic calculus.}\label{tab:sem}
\[\begin{array}{|c|}
\hline 
\mathit{prefix} \quad \quad \alpha \, . \, P \arrow{\alpha} P \\
\mathit{choice} \quad \quad \infr{P_1 \arrow{\alpha} P'_1}{P_1 + P_2
\arrow{\alpha} P'_1} 
\hspace{8mm}
\infr{P_2 \arrow{\alpha} P'_2}{P_1 + P_2 \arrow{\alpha} P'_2} \\
\mathit{recursion} \quad \quad
B \eqdef P \hspace{4mm} \infr{P \arrow{\alpha}{} P'}{B \arrow{\alpha}{} P'} 
\\
\hline
\end{array}\]
\end{table}

\subsection{Interacting agents}

When passing to concurrent processes, we deal with process term instances, called
agents, which represent elements exhibiting the behavior associated to a given 
process term. This separation of concerns between the definition of agents
and of their behavioral pattern is inspired by process algebraic architectural
description languages (see, e.g., \cite{ABC} and the references therein).
The kernel of the semantics of an agent $I$ belonging to the behavioral pattern
defined by process term $P$ is obtained from $P$ by replacing each action
$\alpha$ of $P$ with $I.\alpha$. 
Hence, the semantics $\lsp I \rsp$ of agent $I$ derives from $\lsp P \rsp$ in
the same way.
Then, we say that $I$ is of type $P$, denoted $I : P$, and with the notation
$I.B$ we express that the local behavior of $I$ is given by the process term 
identified by the constant $B$. 
In the following, $\cals$ denotes a finite set of agents $\{I_i : P_i \mid 1 \le
i \le n \}$ such that each agent name $I_i$ is unique.

For notational convenience, from now on, $P,P'\dots$ represent the kernel
of the semantics of agents, hence $P \arrow{I.\alpha} P'$ denotes a transition
performed by agent $I$ from its current local state represented by process term
$P$ to the new local state represented by process term $P'$.
Given a set $\cals$ of agents forming a system, a vector of processes expressing the 
local state of each agent in $\cals$ represents the global state of the system, ranged 
over by $\calp, \calp', \dots$.
Moreover, $\calp[P'/P]$ represents the substitution of $P$ with $P'$ in $\calp$.
Such a notation is not ambiguous as $P,P'$ express the kernel of the semantics of
a uniquely identified agent in $\cals$.

As we will see, the interacting semantics of $\cals$ is given by the parallel 
composition of its constituting agents, the interactions among which are regulated 
by communication rules that depend on community membership and trust information.
In particular, the communication model is based on the following structures:
\begin{itemize}
\item A synchronization set $S \subseteq \mathit{Name}_o \times \mathit{Name}_i$, 
containing pairs of actions denoted syntactically by $a \times b$. Action $a$
represents the output, governing counterpart of the synchronous communication, while 
action $b$ denotes the input, reacting counterpart. Hence, we assume that synchronous 
communication is asymmetric, in the sense that one of the two agents involved governs 
it while the other one reacts.
\item A set of groups of agents (also said set of communities) $\calg \subseteq
2^{\cals}$, such that each group represents a set of agents that can communicate
directly with each other and can share trust opinions. As we will see,
synchronous communication is possible only within the same group, while group 
membership is dynamic.
\item A multiset of trust opinions $\cale$ with support set of type $(\cals,\mathtt{T}
\cup \{?\})_{\cals}$, where $\mathtt{T}$ is a totally ordered trust domain. 
Element $(J,v)_I$ expresses that after a communication between $I$ and $J$, agent $I$ 
has rated the interaction by assigning the score $v$ (the special symbol $?$
means that an occurred interaction has not been rated yet).
We observe that $\cale$ is a multiset, as agent $I$ may be involved in several
different interactions with agent $J$, and some of them could be rated with the same
score. As we will see, trust opinions feed the trust system in order to compute the
trust values $t_{IJ}$, which in turn govern potential synchronous communications
from $I$ to $J$.
\end{itemize}

Intuitively, a trust adaptive system is a set of interacting agents obeying the
communication model described above. 
Therefore, formally, a trust adaptive system is a tuple made of a set of agents 
$\cals$, a synchronization set $S$, a dynamic set of communities $\calg$, and
a dynamic multiset of trust opinions $\cale$ (another parameter, i.e., the trust 
model, is implicit).
The evolution of a trust adaptive system is described by the semantics rules of
Table~\ref{tab:intsem}, which formalize the parallel composition of the agents 
forming the system. More precisely, these rules define the moves (deriving from
autonomous actions and synchronous communications) from configurations to
configurations, where a configuration is defined by the global state of the
system, the synchronization set, the current set of interacting communities, and
the current multiset of trust opinions. Let us explain intuitively such rules.

\begin{table}[tbh]
\caption{Semantics rules for parallel composition.}\label{tab:intsem}
\[\begin{array}{|c|}
\hline 
\infr{P \in \calp \quad P \arrow{I.\tau} P'}
{(\calp,S,\calg,\cale) \arrow{I.\tau} (\calp[P'/P],S,\calg,\cale)} 
\\[2mm] 
\infr{P \in \calp \quad G \in \calg \quad P \arrow{I.\mathit{ent}(G)} P'}
{(\calp,S,\calg,\cale) \arrow{I.\tau} (\calp[P'/P],S,\calg[G \cup \{I\} / G],\cale)}
\\[2mm]
\infr{P \in \calp \quad G \in \calg \wedge I \in G \quad P \arrow{I.\mathit{esc}(G)} P'}
{(\calp,S,\calg,\cale) \arrow{I.\tau} (\calp[P'/P],S,\calg[G \backslash \{I\} / G],\cale)}
\\[2mm]
\infr{P_1,P_2 \in \calp, P_1 \not= P_2 \quad a \times b \in S \quad G \in \calg \wedge I,J \in G \quad 
P_1 \arrow{I.a} P'_1 \quad P_2 \arrow{J.b} P'_2 \quad a \in H \,\wedge\, t_{IJ} \ge \mathit{th}_{I}}
{(\calp,S,\calg,\cale) \arrow{I.a \times J.b}
(\calp[P'_1/P_1,P'_2/P_2],S,\calg,\cale \cup \{\!\!|(J,?)_I|\!\!\} \cup \{\!\!|(I,?)_J|\!\!\} )}
\\[2mm]
\infr{P_1,P_2 \in \calp, P_1 \not= P_2 \quad a \times b \in S \quad G \in \calg \wedge I,J \in G \quad 
P_1 \arrow{I.a} P'_1 \quad P_2 \arrow{J.b} P'_2 \quad a \in L \,\wedge\, t_{IJ} < \mathit{th}_{I}}
{(\calp,S,\calg,\cale) \arrow{I.a \times J.b} 
(\calp[P'_1/P_1,P'_2/P_2],S,\calg,\cale \cup \{\!\!|(J,?)_I|\!\!\} \cup \{\!\!|(I,?)_J|\!\!\} )}
\\[2mm]
\infr{P_1,P_2 \in \calp, P_1 \not= P_2 \quad a \times b \in S \quad G \in \calg \wedge I,J \in G \quad 
P_1 \arrow{I.a} P'_1 \quad P_2 \arrow{J.b} P'_2 \quad a \not\in \{H \cup L\}}
{(\calp,S,\calg,\cale) \arrow{I.a \times J.b} 
(\calp[P'_1/P_1,P'_2/P_2],S,\calg,\cale)} 
\\[2mm]
\infr{P \in \calp \quad G \in \calg \wedge I,J \in G \quad (J,?)_I \in \cale \quad
P \arrow{I.\mathit{obs}(v)} P'}
{(\calp,S,\calg,\cale) \arrow{I.\tau} (\calp[P'/P],S,\calg,\cale \backslash
\{\!\!|(J,?)_I|\!\!\} \uplus \{\!\!|(J,v)_I |\!\!\})}
\\
\hline
\end{array}\]
\end{table}

The first rule refers to the internal action $\tau$, which is performed autonomously by 
each agent.
Then, we have two additional, internal actions that can be performed autonomously
by each agent, which we add to the syntax of the basic calculus:
\[ 
\mathit{ent}(G) \mid \mathit{esc}(G)
\]
where $G \in \calg$. Such actions concern the membership to communities.  In
particular, action $\mathit{ent}(G)$ allows an agent to join the group $G$ of
agents (notice that $G$ is replaced by $G \cup \{I\}$, where $I$ is the agent
joining the group). Action $\mathit{esc}(G)$ allows an agent to leave the group
$G$ of agents (notice that $G$ is replaced by $G \backslash \{I\}$, where $I$ is
the agent leaving the group).  We point out that groups are used to dynamically
confine the sets of agents that can interact directly through synchronous
communication and within which trust based information can be shared. Hence,
such sets represent the communities referenced by an agent in a given instant of
time in order to obtain trust recommendations.

The following three rules formalize the trust-based synchronous communication
between two different agents.
Based on the communication model previously described, an interaction from $I$,
offering output $a$, to $J$, reacting with input $b$, is possible if two
conditions hold:
\begin{itemize}
\item $a \times b$ belongs to the synchronization set $S$;
\item there exists a community of which both $I$ and $J$ are members.
\end{itemize}
Moreover, the communication from $I$ to $J$ may depend on the
trust of $I$ towards $J$.
Inspired by the noninterference approach to information flow
analysis~\cite{GM82}, all the actions involved in trust-based communications are
classified into two disjoint sets, $H$ and $L$, denoting high-level and
low-level actions, such that: 
\begin{itemize}
\item $(H \cup L) \subseteq \mathit{Name}$;
\item for each $a \times b \in S$ it holds that $a \in H$ if and only if $b \in
H$ and $a \in L$ if and only if $b \in L$.
\end{itemize}
If agent $I$ offers output $a \in H$, then the potential reacting counterpart 
must satisfy the trust-based selection policy based on the trust threshold 
$\mathit{th}_I$. A typical high-level action is the service request sent by an
agent $I$ to another agent $J$, which is chosen as a trusted partner. 
Notice that since the communication model is asymmetric, then the trust-based 
condition is applied only by the agent offering the output action, which governs 
the interaction.
On the contrary, if agent $I$ offers output $a \in L$, then an interaction
through $a$ is possible only if the trust-based selection policy based on the trust
threshold $\mathit{th}_I$ is not satisfied by the counterpart. A typical low-level 
action is the denial of service delivery that is sent by an agent $I$ to another 
agent $J$, who previously sent a service request to $I$ that cannot be accepted as 
$J$ is not trusted enough by $I$.
If $a \not\in \{H \cup L\}$, then every interaction involving $a$ does not rely 
on trust-based requirements. 
The trust-based selection policy enabling a trusted interaction from $I$ to $J$ is 
$t_{IJ} \ge \mathit{th}_{I}$, where $t_{IJ}$ is the trust of $I$ towards $J$ as 
estimated by the trust model, which relies on the set of basic parameters collected 
during the system execution. 
Hence, its calculation strictly depends on the chosen trust model and does not 
affect the definition of the semantics for interacting processes. 
As discussed, $\mathit{t}_{IJ}$ may be based on several different methods~\cite{KSG,ZH,ZLH}, 
an example of which will be given in the following.
Whenever a trust-based communication occurs, then a feedback, in the form of a score $v$, 
could be provided by each of the two parties to rate the level of satisfaction in the
interaction with the other party. To keep track of such a possibility, terms
$(J,?)_I$ and $(I,?)_J$ are added to the set $\cale$ of local opinions. 
The former denotes that $I$ can rate an interaction with $J$, and vice versa for
the latter. This evaluation may occur later on during system execution. 
Hence, to report the feedback, we add to the syntax of the basic calculus the
special internal action $\mathit{obs}(v)$, where $v \in \mathtt{T}$, which allows the
agent executing it to rate a trust-based interaction previously conducted with a
known agent, see the last semantic rule.
Notice that the effect of such an action is to replace the symbol $?$ in
$(J,?)_I$ with the score $v$.

\textbf{Well-formedness.}
The placeholder $(J,?)_I$ is added to the multiset $\cale$ through the union 
operator $\cup$~\footnote{Multiset union is defined as the multiset such that
each element has the maximal multiplicity it has in either multisets.}. As a
consequence, it can occur in $\cale$ with multiplicity $1$ at most. 
A score assigned to an interaction between $I$ and $J$ refers to the last 
of the unrated interactions among them. If other, older, unrated interactions among 
them exist, they lose the possibility to be rated. In this way, we can model the 
situation in which no feedback is reported, either because it is not needed or 
when the user is not stimulated to provide trust rates.
Whenever the placeholder $(J,?)_I$ is removed, an element of the form
$(J,v)_I$ is added to $\cale$ through the multiset sum operator
$\uplus$~\footnote{Multiset sum is defined as the multiset such that each
element has the sum of the multiplicities it has in both multisets.}, meaning 
that such an element may occur in $\cale$ with multiplicity greater than $1$.
Notice that, if two different placeholders $(J,?)_I$ and $(J',?)_{I}$ occur
in $\cale$, then the execution of transition $I.\mathit{obs}(v)$ assigns score $v$
either to $J$ or to $J'$, nondeterministically. 
Such a situation is avoided if the feedback is reported before the execution of a 
new interaction with another agent, as typical in most trust-based systems, in which 
case we say that the system is well-defined.

As far as the feedback mechanism is concerned, the last rule of the semantics
expresses the correct behavior of an agent rating a real interaction, as
expected by any trust system. However, such an assumption is a limitation with
respect to the modeling of malicious behaviors, which would require an improper
use of the action $\mathit{obs}$.  With the aim of modeling fake trust reports
and, therefore, false recommendations, we add a new special internal action and the
following rule for pushing fictitious opinions:
\[
\infr{P \in \calp \quad G \in \calg \wedge I,J \in G \quad
P \longarrow{I.\mathit{fake\_obs}(J,v)} P'}
{(\calp,S,\calg,\cale) \arrow{I.\tau} (\calp[P'/P],S,\calg, 
\cale \uplus \{\!\!|(J,v)_I |\!\!\})}
\]
which allows any agent to rate the other agents of the community without any
restriction.
Going back to the list of attacks discussed in Section~\ref{sec:intro}, we
observe that they can be modeled by using actions $\mathit{obs}(v)$ and 
$\mathit{fake\_obs}(J,v)$. Moreover, the adaptive community-based communication
policy is useful to model sybil and white-washing attacks.

The formal semantics of interacting agents is expressed in terms of an extension of
LTSs.

\begin{definition} 
Given a set of trust predicates $\mathbb{TP}$ and a set of names $\mathbb{N}$, a
trust labeled transition system (TLTS) is a tuple $(Q,q_0,L,R,T)$ where:
\begin{itemize}
\item $(Q,q_0,L,R)$ is a LTS. 
\item $T: Q \rightarrow 2^\mathbb{TP} \times 2^\mathbb{N}$ is a labeling
function.
\end{itemize}
\end{definition}

In our framework, $\mathbb{TP}$ is of the same type as $\cale$, while
$\mathbb{N}$ is the set of agent names.
Then, the semantics of a trust adaptive system described by the tuple 
$(\cals, S, \calg, \cale)$, where $\cals$ contains agents $I_i,\ 1 \le i \le n$,
$S$ is the synchronization set, $\calg$ is the initial set of communities, and 
$\cale$ is the initial multiset of trust opinions, is the smallest TLTS satisfying
the following conditions:
\begin{itemize}
\item Each global state $q \in Q$ is a $n$-length vector of process terms modeling 
the local behavior of the agents $I_i,\ 1 \le i \le n$, such that the initial global 
state $q_0$ is associated to the vector modeling the initial local state of each
agent.
\item $L = \{I.\tau \,\mid\, I \in \cals \} \cup \{I.a \times J.b \,\mid\, I,J
\in \cals \wedge a \times b \in S\}$.
\item The transitions in $R$ and the labelings of $T$ are obtained through the 
application of the operational semantics rules of Table~\ref{tab:intsem}, with
the labels of $q_0$ determined by $\calg$ and $\cale$.
\end{itemize}
Typically, $\cale = \emptyset$ in $q_0$. The assumption concerning the emptiness
of $\cale$ in the initial state can be changed according to the trust model. In
some case (see, e.g., \cite{KSG}), in fact, a priori estimations of trust are
assigned to agents that are known to be trustworthy in a community, e.g., as
they are among the founders of the community. Hence, pre-trusted agents can be
modeled by setting adequately $\cale$ in the initial state.

\section{Two examples}\label{sect:example}

In this section, we sketch the formal modeling of two real-world systems using 
the trust models of~\cite{ZLH} and~\cite{KSG}, in which local trust
deriving from direct experiences is calculated by counting the number of
positive and negative experiences. 
Hence, it is sufficient to assume that the feedback
reported through action $\mathit{obs}$ is either $1$ or $-1$, respectively, thus
implying $\mathtt{T} = \intns$. 

First, let us consider a system using the trust model proposed in~\cite{ZLH}. 
In the following, we illustrate the main aspects related to the computation of
the trust value $t_{IJ}$ without going into the details of the algebraic 
specifications expressing the agents behavior.
The system includes behavioral patterns for the following categories: nodes
consuming services (type \textit{Cons}), nodes delivering services (type
\textit{Prod}), and nodes governing clubs (type \textit{CDSR}).
Each club is defined as a group including one agent of type CDSR, some consumer, 
and several producers offering the service that characterizes the club.
For instance, given two fixed clubs $G_1$ and $G_2$, the process term \textit{Cons}
could be defined as follows (the summation symbol $\sum$ is used to generalize
the choice operator):
\[\begin{array}{ll}
\mathit{Cons} \eqdef & \sum_{i \in \{1,2\}} \tau . 
\mathit{send\_request_i} . ( 
\\
& \sum_{\{j \in G_i \}} \mathit{receive\_service_{ij}} .
(\mathit{obs(1)}.\mathit{Cons} + \mathit{obs(-1)} . \mathit{Cons}) 
+ \mathit{receive\_denial_{ij}} . \mathit{obs(-1)} . \mathit{Cons})
\end{array}\]
where output $\mathit{send\_request_i} \in H$, input
$\mathit{receive\_service_{ij}} \in H$, and input $\mathit{receive\_denial_{ij}} 
\in L$. We assume that the synchronization set enables a communication through
$\mathit{send\_request_i}$ and a corresponding input, say
$\mathit{receive\_request_i}$, which is offered by every producer $j$ belonging
to $G_i$. Notice that the choice of the specific producer $j$ is nondeterministic 
among the agents trusted by the consumer, which proposes the request to all the
agents of group $G_i$. Such an interaction is not rated by the consumer.
Afterwards, through adequate synchronizations between the consumer and the
responding producer, either the consumer receives the service, and then rates the
interaction nondeterministically, or the producer refuses the request, and in such 
a case the consumer rates negatively the failure. The choice between the two
events is deterministic and based on the trust of the chosen producer towards
the consumer.
We point out that the feedback, reported through action $\mathit{obs}$, is assigned 
to the unique producer interacting with the consumer in a fully transparent way by 
virtue of the semantics rules of Table~\ref{tab:intsem}.  

All the interactions governed by trust are based on the following encoding of
the trust model of~\cite{ZLH}.
Given agent $k$ in the club $Y$, Equation~\ref{eq:trustinclub} is estimated by
setting parameters $p$ and $n$ as follows:
\[p = \sum_{j \in Y, j \not= k} \mathit{mul}((k,1)_j)\]
where $\mathit{mul}(e)$ denotes the multiplicity of term $e$ in $\cale$.
The estimation of parameter $n$ is analogous by replacing $1$ with $-1$ in the
definition above.
On the other hand, given clubs $X$ and $Y$, Equation~\ref{eq:clubtrust} is
estimated as follows:
\[p = \sum_{i \in X, j \in Y} \mathit{mul}((j,1)_i)\]
and similarly in the case of parameter $n$.
Given such a model, any trust-based communication enabled in a global state $q$
of the TLTS representing the current system behavior, depends on the labeling
$T(q)$. Notice that, in order to allow agents of different clubs 
to interact directly, the system includes ad-hoc groups of the form $\{i,j\}$ 
enabling the communication between $i$ and $j$. On the other hand, the communication 
is allowed (or not), depending on the trust $t_{ij}$ computed as shown above.

As another example, let us consider the encoding of EigenTrust~\cite{KSG} in our 
framework.
First, observe that the local trust from $I$ to $J$ is given by 
$s_{IJ} = \mathit{mul}((J,1)_I) - \mathit{mul}((J,-1)_I)$.
Then, $c_{IJ}$ is obtained through the normalization function defined in~\cite{KSG}.
Hence, the formula used to compute $t_{IJ}$ is
$\mathit{trust}^{\{I,J\}}_{IJ}$, where:
\[ \mathit{trust}^{S}_{IJ} = c_{IJ} + 
\sum_{G\,\mathit{s.t.}\, I \in G} \
\sum_{K \in G, K \not\in S} c_{IK} \cdot 
\mathit{trust}^{\{K\} \,\cup\, S}_{KJ}.
\]

\section{Model checking trust properties}\label{sect:mc}

The formal framework proposed in this paper can be used as a basis for the
verification of distributed trust systems. For this purpose, in~\cite{Ald15}, a
model checking based approach is defined that relies on a trust temporal logic,
called TTL, which is defined for the verification of TLTS-like models and, e.g., 
can be mapped to the logic UCTL~\cite{BFGM}.
Here, we specify the atomic statements of such a logic, which depend on the
representation of trust information in our calculus, while the logical and temporal
operators can be found in~\cite{Ald15}. Similarly as for other logics merging 
action/state based predicates, atomic formulas include actions labeling TLTS 
transitions and state-based trust predicates:
\[ \i \mid w \ge k \]
where the domain of variable $\i$ is the labels set $L$ of the TLTS, $k \in \mathtt{T}$, 
and $w$ is a trust variable, which can be equal to: 
\begin{itemize}
\item $t_{IJ}$, i.e., the trust of $I$ towards $J$ as computed by the
trust system;
\item $\mathit{tf}_{IJ} = f \{\!\!| v \,|\, (J,v)_I \in \cale |\!\!\} $,
where function $\mathit{f} : 2^{\mathtt{T}} \rightarrow \mathtt{T}$ is taken
from a set $\mathit{TF}$ of associative and commutative functions, like, e.g., sum, 
min, and count, provided that $\mathtt{T} = \intns$.
\end{itemize}
Therefore, an atomic statement is a predicate about either the trust between two
agents as computed by the trust system, or the set of local, direct experiences between 
them.
In this framework, trust temporal properties can be modeled and verified, like, e.g., 
``Can $n$ malicious agents provide false feedback in order to compromise the
reputation of a honest agent?'', or ``Can an agent trust another agent without
sufficient, positive, direct observations?'', thus making it possible the 
validation of a system against the attacks mentioned in Section~\ref{sec:intro}.

\section{Conclusion and future work}\label{sect:conc}

The formal modeling approach proposed in this paper joins the specification of
distributed systems relying on an adaptive and flexible communication model with 
the specification of the trust model governing the interactions among concurrent
processes. These two modeling frameworks are defined separately, as the mutual
interaction between them is managed transparently at the level of the semantics
of parallel composition.

As work in progress, we mention that the multiset of trust values storing the
feedback about direct interactions can be enriched with additional information,
such as, e.g., the age of each feedback. This can be done in order to weight the 
contribution of an experience depending on the time elapsed from the related 
interaction.

The information expressed by the trust infrastructure is employed to make
the model quantitative, in a sense, without adding \textit{numbers} to the
behavioral specification of the agents. Such quantitative information can be
used to solve nondeterminism in several different ways.
For instance, the possibilistic choice among alternative trust-based
communications from agent $i$ to a set of trusted agents $X$ can be made either
probabilistic, by using as weights the trust of $i$ towards each agent $j$ in
$X$, or prioritized, by using the same trust values, or else a combination of
the two policies can be applied. Details about the extension of the TLTS model
that is obtained in such a way, which encompasses both nondeterminism and
probabilities, can be found in~\cite{Ald15}.

Finally, as future work, it would be worthwhile to parameterize (without any
substantial human intervention) the model checking based verification with
respect to the different classes of attacks described in Section~\ref{sec:intro}.

\bibliographystyle{eptcs}
\bibliography{biblio}

\end{document}

%% file: commands.tex
\newcommand{\cale}
        {\mathcal{E}}

\newcommand{\calg}
        {\mathcal{G}}

\newcommand{\cals}
        {\mathcal{S}}

\newcommand{\calp}
        {\mathcal{P}}

\newcommand{\intns}
	{\mathbb{Z}}

\newcommand{\nil}
  {\underline 0}

\newcommand{\eqdef}
  {\stackrel{\mathrm{def}}{=}}

\newcommand{\infr}[2]
	{\renewcommand{\arraystretch}{1.5}
	\begin{array}{c}
	#1\\
	\hline
	#2
	\end{array}}

\newcommand{\arrow}[1]
        {\, {\auxarrow\limits^{#1}} \,}
\newcommand{\auxarrow}
	{\mathop{- \!\!\!\! \longrightarrow}}

\newcommand{\longarrow}[1]
        {\, {\auxlongarrow\limits^{#1}} \,}
\newcommand{\auxlongarrow}
	{\mathop{- \hspace{-0.2cm} - \hspace{-0.2cm} - \hspace{-0.2cm}
	- \hspace{-0.2cm} - \hspace{-0.2cm} - \hspace{-0.2cm}
	- \hspace{-0.2cm} - \hspace{-0.2cm} - \hspace{-0.2cm}
	- \hspace{-0.2cm} - \hspace{-0.3cm} \longrightarrow}}

\newcommand{\lsp}
	{[ \! [}
\newcommand{\rsp}
	{] \! ]}



%% file: environments.tex
\newtheorem{new_theorem}
	{Theorem}[section]

\newtheorem{new_definition}
	[new_theorem]{Definition}

\newtheorem{new_remark}
	[new_theorem]{Remark}

\newtheorem{new_example}
	[new_theorem]{Example}

\newtheorem{new_lemma}
	[new_theorem]{Lemma}

\newtheorem{new_proposition}
	[new_theorem]{Proposition}

\newtheorem{new_corollary}
	[new_theorem]{Corollary}

\newenvironment{definition}
	{\begin{new_definition}\rm}
	{\end{new_definition}}

\newenvironment{example}
	{\begin{new_example}\rm}
	{\end{new_example}}

